\documentclass[prl]{revtex4}
\usepackage{amssymb}
\usepackage{makeidx}
\usepackage{graphicx}

\input{tcilatex}

\begin{document}

\title{The Raman response of double wall carbon nanotubes}
\author{F. Simon, R. Pfeiffer, C. Kramberger, M. Holzweber, and H. Kuzmany}
\affiliation{Institute of Materials Physics, University of Vienna, A-1090 Vienna,
Strudlhofgasse 4., Austria }

\begin{abstract}

Raman spectroscopy on carbon nanotubes (CNT) yields a rich variety of
information owing to the close interplay between electronic and vibrational
properties. In this paper, we review the properties of double wall carbon
nanotubes (DWCNTs). In particular, it is shown that SWCNT encapsulating C$_{60}$, so-called peapods, are transformed into DWCNTs when subject to a
high temperature treatment. The inner tubes are grown in a catalyst free
environment and do not suffer from impurities or defects that are usually
encountered for as-grown SWCNTs or DWCNTs. As a consequence, the inner tubes
are grown with a high degree of perfection as deduced from the unusually
narrow radial breathing mode (RBM) lines. This apostrophizes the interior of
the SWCNTs as a nano-clean room. The mechanism of the inner nanotube
production from C$_{60}$ is discussed. We also report recent studies aimed
at the simplification and industrial scaling up of the DWCNT production
process utilizing a low temperature peapod synthesis method. A splitting of
the RBMs of inner tubes is observed. This is related to the interaction
between the two shells of the DWCNTs as the same inner tube type can be
encapsulated in different outer ones. The sharp appearance of the inner tube
RBMs allows a reliable assignment of the tube modes to (n,m) indexes and
thus provides a precise determination of the relation between the tube
diameter and the RBM frequencies.
\end{abstract}

\maketitle

\section{Introduction}

Carbon nanotubes have been in the forefront of the nanomaterial research
since their discovery \cite{IijimaNAT}. They are not only fundamentally
interesting materials due to their appealing one-dimensional structure but
several applications have been envisaged. Some of them have already been
established such as scanning probe-heads \cite{HafnerNAT} or field emission
devices \cite{ZhouAPL}\cite{Obraztsov}. There is an active ongoing work in
these fields to exploit the properties of these materials better and to
improve the device qualities. Furthermore, high expectations are related to
their applications as building elements of electronics, composite
reinforcing materials and many more.

Carbon nanotubes can be represented as rolled up graphene sheets, i.e.
single layers of graphite. Depending on the number of coaxial carbon
nanotubes, they are usually classified into multi-wall carbon nanotubes
(MWCNTs) and single wall carbon nanotubes (SWCNTs). Some general
considerations have been clearified in the last 13 years of nanomaterial
research related to these structures. MWCNTs are more homogeneous in their
physical properties as the large number of coaxial tubes smears out
individual tube properties. This makes them suitable candidates for
applications where their nanometer size and the conducting properties can be
exploited. In contrast, SWCNT materials are grown as an ensemble of weakly
interacting tubes with different diameters. The physical properties of
similar diameter SWCNTs can change dramatically as the electronic structure
is very sensitive on the rolling-up direction, the so-called chiral vector.
Depending on the chiral vector, SWCNTs can be metallic or semiconducting 
\cite{Dresselhaus}. This provides a richer range of physical phenomena as
compared to the MWCNTs, however significantly limits the range of
applications. To date, neither the directed growth nor the controlled
selection of SWCNTs with a well defined chiral vector has been performed
successfully. Thus, their broad applicability is still awaiting.
Correspondingly, current reserach is focused on the post-synthesis
separation of SWCNTs with a narrow range of chiralities \cite%
{ChattopadhyayJACS}\cite{KrupkeSCI}\cite{RinzlerNL}\cite{StranoSCI} or on
methods which yield information that are specific to SWCNTs with different
chiralities. An example for the latter is the observation of chirality
selective band-gap fluorescence in semiconducting SWCNTs \cite{WeismanSCI}.

A more recently discovered third class of CNTs are double-wall carbon
nanotubes (DWCNTs). DWCNTs were first observed to form under intensive
electron radiation \cite{LuzziCPL1999} in a high resolution transmission
electron microscope from C$_{60}$ encapsulated in SWCNTs, so-called peapods 
\cite{SmithNAT}. Following the synthesis of C$_{60}$ peapods in macroscopic
amounts \cite{KatauraSM}, bulk quantities of the DWCNT material are
available using a high temperature annealing method \cite{BandowCPL}.
Alternatively, DWCNTs can be produced with usual synthesis methods such as
arc-discharge \cite{HutchisonCAR} or CVD \cite{ChengCPL} under special
conditions. According to the number of shells, DWCNTs are between SWCNTs and
MWCNTs. Thus, one expects that DWCNTs may provide a material where improved
mechanical stability as compared to SWCNTs coexists with the rich variety of
electronic properties of SWCNTs. There are, of course, a number of yet
unanswered questions e.g. if the outer tube properties are unaffected by the
presence of the inner tube or if the commensurability of the tube structures
plays a role. These questions should be answered before the successful
application of these materials.

In this contribution, we review Raman studies of DWCNTs. We show that the
study of inner tubes, in particular those from C$_{60}$ peapod based DWCNTs
provides some unique insight into the physics of SWCNTs. Such studies enabled
the observation of unprecedently sharp Raman modes, which evidence that the
inside grown SWCNTs are higly perfect mainly due to the catalyst free
nano-clean room interior of outer SWCNT reactor tubes. The sharp Raman
features of inner tube Raman radial breathing modes (RBMs) enable the
indexing of chiral vectors thus providing a simple alternative to the
band-gap fluorescence method.

This review is organised as follows: we describe the experimental methods
and the sample preparation that are used for the current study. We compare
the properties of DWCNTs grown with different methods. We show that the C$%
_{60}$ peapod based DWCNTs have unique properties which underline the
nano-clean room conditions encountered in the inside of SWCNTs. We describe
a novel method for the preparation of the C$_{60}$ peapod precursor material
that enables the large scale productions of DWCNTs. We present a detailed
investigation of the electronic structure of the small diameter inner tubes.
We also present the chiral vector assignment to such tubes thus refining the
empirical parameters of the relation between the RBM frequencies and the
tube diameters.

\section{Experimental}

SWCNT starting materials for the production of DWCNTs described herein were
prepared by the laser ablation method. Their diameters were controlled in
order to obtain efficient C$_{60}$ encapsulation that results in high yield
of inner nanotubes. The values of $d_{\mathrm{L}}=$ 1.39\ nm, $\sigma _{\mathrm{L}%
}$ = 0.1 nm were obtained for the mean diameter and the variance of the
distribution for the different samples using a large number of exciting
laser energies following Ref. \cite{KuzmanyEPJB}. The SWCNT materials were
purified following Ref. \cite{KatauraSM}. Peapod samples were prepeared by
annealing SWCNTs with C$_{60}$ in a quartz ampoule following Ref. \cite%
{KatauraSM}. The peapod filling fraction was close to 100\% as evidenced
previously on similar samples using Electron Energy Loss Spectroscopy (EELS) 
\cite{LiuPRB}. The peapod materials were transformed to DWCNTs using the
high temperature annealing method of Ref. \cite{BandowCPL}. The samples in
the form of bucky-paper are kept in dynamic vacuum and on a copper tip
attached to a cryostat, which allows temperature variation in the 20-600 K
temperature range. The Raman experiments were performed in a 180 degree
backscattering geometry. A He/Ne, an Ar/Kr mixed gas, and a tunable
Ti:sapphire laser pumped by an Ar laser were used for the excitation at 30
different laser lines. Multi frequency Raman spectroscopy was performed on a
Dilor xy triple axis spectrometer in the 1.64-2.54 eV (755-488 nm) energy
range and in a Bruker FT-Raman spectrometer for the 1.16 eV (1064 nm)
excitation at 90 K. We operated the Dilor spectrometer in two modes, high
and normal resolution. The high resolution uses the additive mode of the
spectrometer and the spectral resoluton as determined from the FWHM of the
elastically scattered light was 0.4-0.7 cm$^{-1}$ going from red to blue
excitation. Similarly, spectral resolution in the normal mode was 1-2 cm$%
^{-1}$ depending on the laser line. Measurements with the FT-Raman
spectrometer were recorded with 1 cm$^{-1}$ resolution. Raman shifts were
accurately calibrated against a series of spectral calibration lamps. 
\textit{Ab initio} calculations were performed with the Vienna Ab Initio
Simulation Package (VASP) \cite{KressePRB}.

\section{Results and discussion}

\subsection{DWCNT synthesis}

\begin{figure}[tbp]
\centering \includegraphics[width=0.6\textwidth]{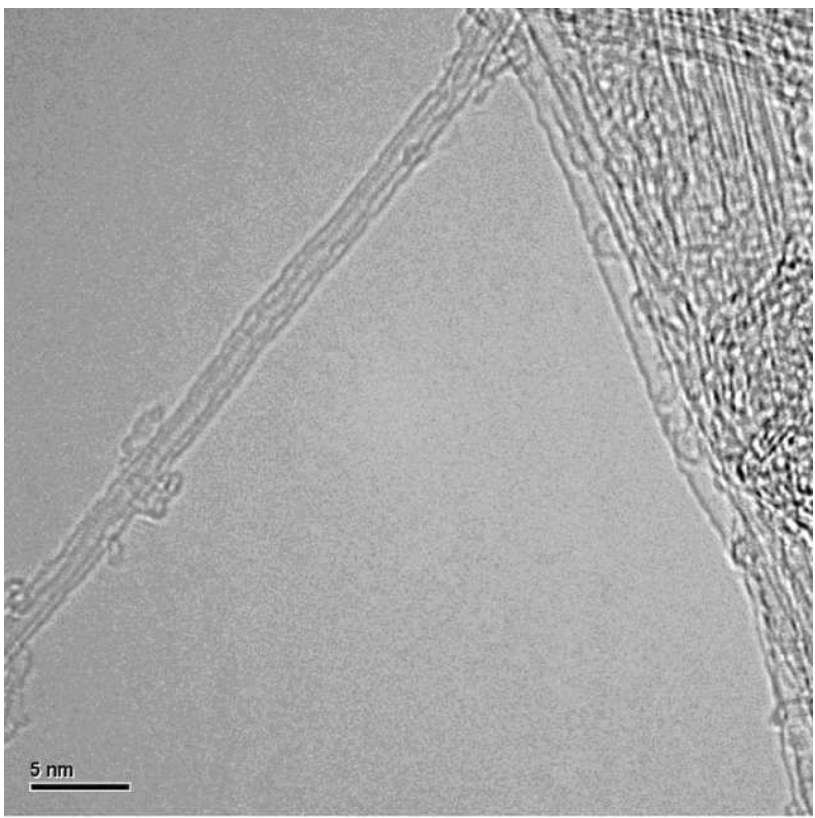}
\caption{High-resolution TEM micrograph of a C$_{60}$ peapod based DWCNT
sample.}
\label{TEM_DWNT}
\end{figure}

Double-wall carbon nanotubes can be classified into two groups depending on
the method used for their production. DWCNT samples produced with usual
preparation methods \cite{HutchisonCAR} have less controllable parameters
such as their diameter distributions and, as shown below, their quality is
inferior compared to the C$_{60}$ peapod based DWCNTs. Following the
discovery of DWCNTs from C$_{60}$ peapods under intensive electron
irradiation \cite{LuzziCPL1999}, it was demonstrated using HR-TEM that a 1200%
$%
%TCIMACRO{\U{b0}}%
%BeginExpansion
{{}^\circ}%
%EndExpansion
$C heat treatment can also efficiently produce the inner nanotubes based on
the C$_{60}$ peapods \cite{LuzziCPL2000}. The first characterization of bulk
amounts of DWCNTs produced by this synthesis method was performed using
Raman spectroscopy by Bandow et al. \cite{BandowCPL}.

\begin{figure}[tbp]
\centering \includegraphics[width=0.6\textwidth]{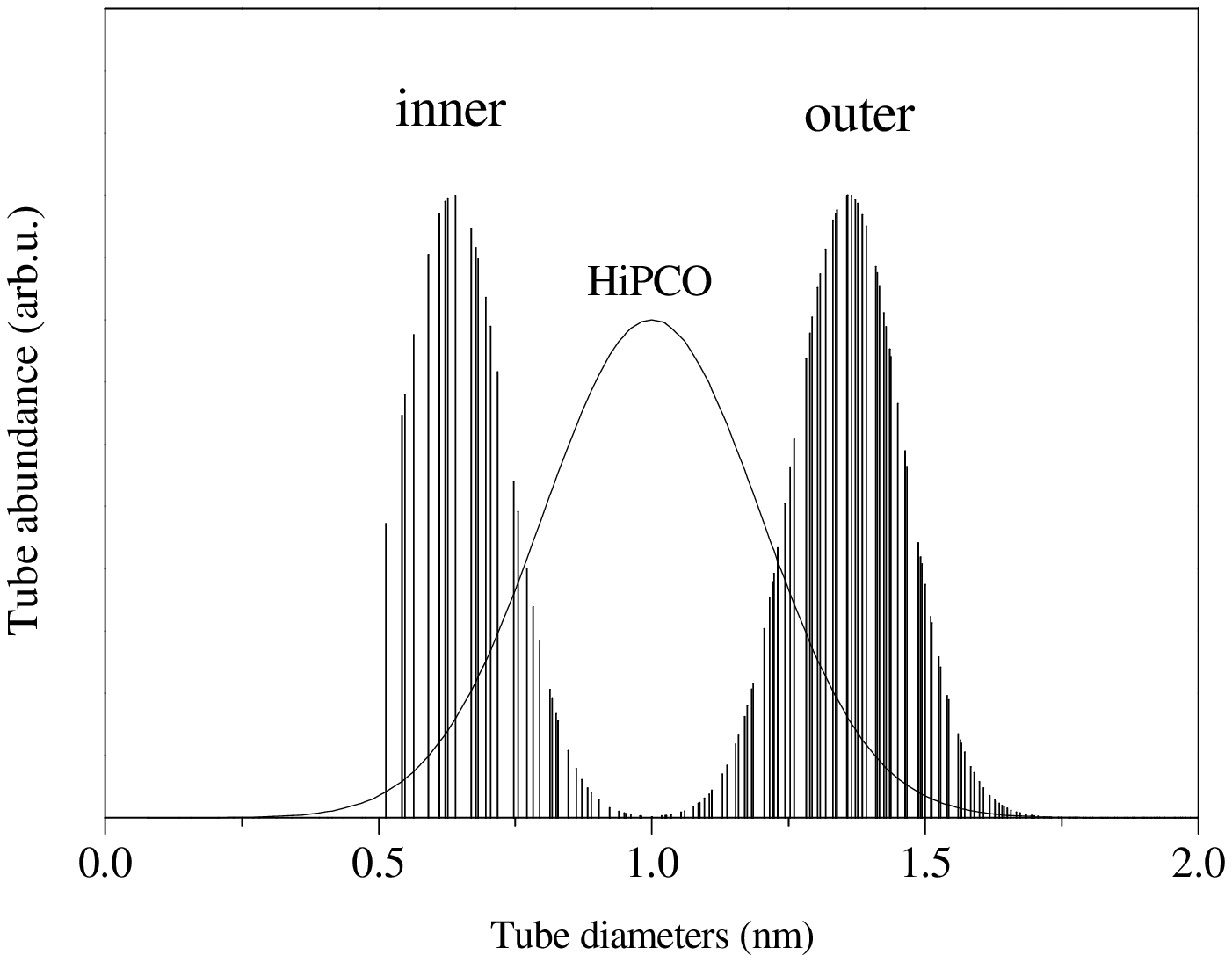}
\caption{Schematic diameter distribution of geometrically allowed tubes in
the DWCNTs of the current study. The envelope of the approximate diameter
distibution in a HiPco sample is shown for comparison.}
\label{DWNT_diams}
\end{figure}

In Fig. \ref{TEM_DWNT} we show a typical HR-TEM micrograph of C$_{60}$
peapod based DWCNT. The difference in the outer and inner tube diameter is
thought to be close to twice the van der Waals distance of graphite, 0.335
nm. Indeed, X-ray studies have indeed shown that the inner and outer tube
diameter difference is only slightly larger, 0.72$\pm $0.02 nm \cite{AbePRB}%
. The corresponding situation for a DWCNT sample with outer tube distibution
centered at the (10,10) tube is depicted in Fig. \ref{DWNT_diams}. The
cut-off observed at small diameter inner tubes is given by the smallest
outer tube diameter of $1.2$ nm which allows the precursor C$_{60}$ to enter 
\cite{ZerbettoJPC}\cite{TomanekPRL}\cite{RochefortCM}.

\begin{figure}[tbp]
\centering \includegraphics[width=0.6\textwidth]{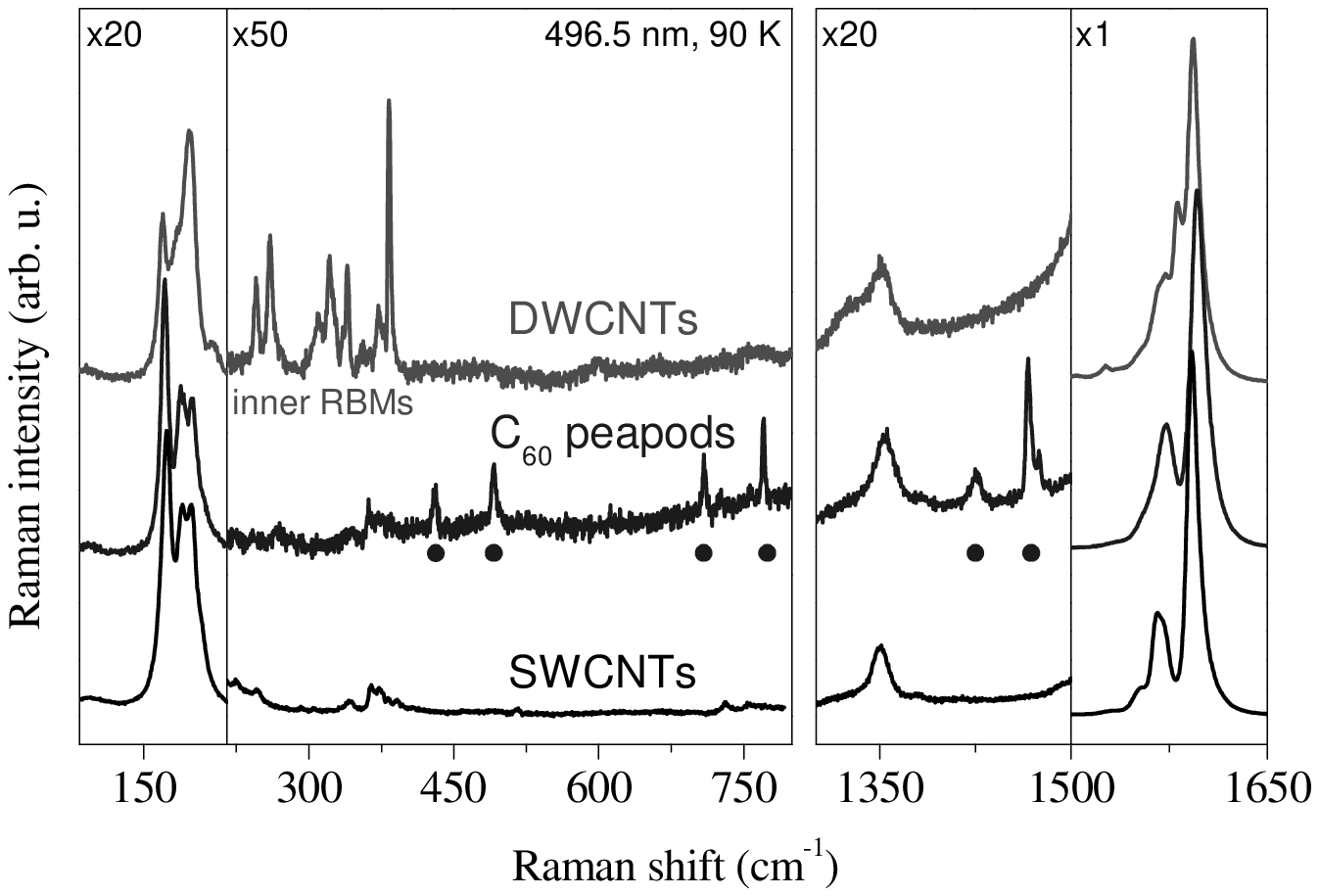}
\caption{Raman spectra of pristine SWCNT, C$_{60}$ peapod and C$_{60}$
peapod based DWCNT samples for $\protect\lambda$=497 nm laser excitation and
90 K. Bullets mark the positions of the modes of encapsulated C$_{60}$.}
\label{DWNT_transf}
\end{figure}

The transformation from C$_{60}$ peapods to DWCNTs can be conveniently
followed by Raman spectroscopy. In Fig. \ref{DWNT_transf} we compare the
Raman spectra of pristine SWCNTs, C$_{60}$@SWCNT peapods, and DWCNTs based
on the peapod material. The emergence of additional vibrational modes in the
200-450 cm$^{-1}$ spectral range is evident in Fig. \ref{DWNT_transf} for
the DWCNT material as compared to the pristine and peapod materials. This
emergence is accompanied by the dissappearance of the modes of the
encapsulated fullerene as denoted by circles in Fig. \ref{DWNT_transf}. The
extra lines in the 200-450 cm$^{-1}$ spectral range were identified as the
radial breathing modes of the smaller diameter inner shell tubes. The
dissappearance of the fullerene peaks is evidence that the encapsulated C$%
_{60}$ serves as carbon source for the internal tube formation \cite%
{BandowCPL}\cite{PfeifferPRL}.

\begin{figure}[tbp]
\centering \includegraphics[width=0.6\textwidth]{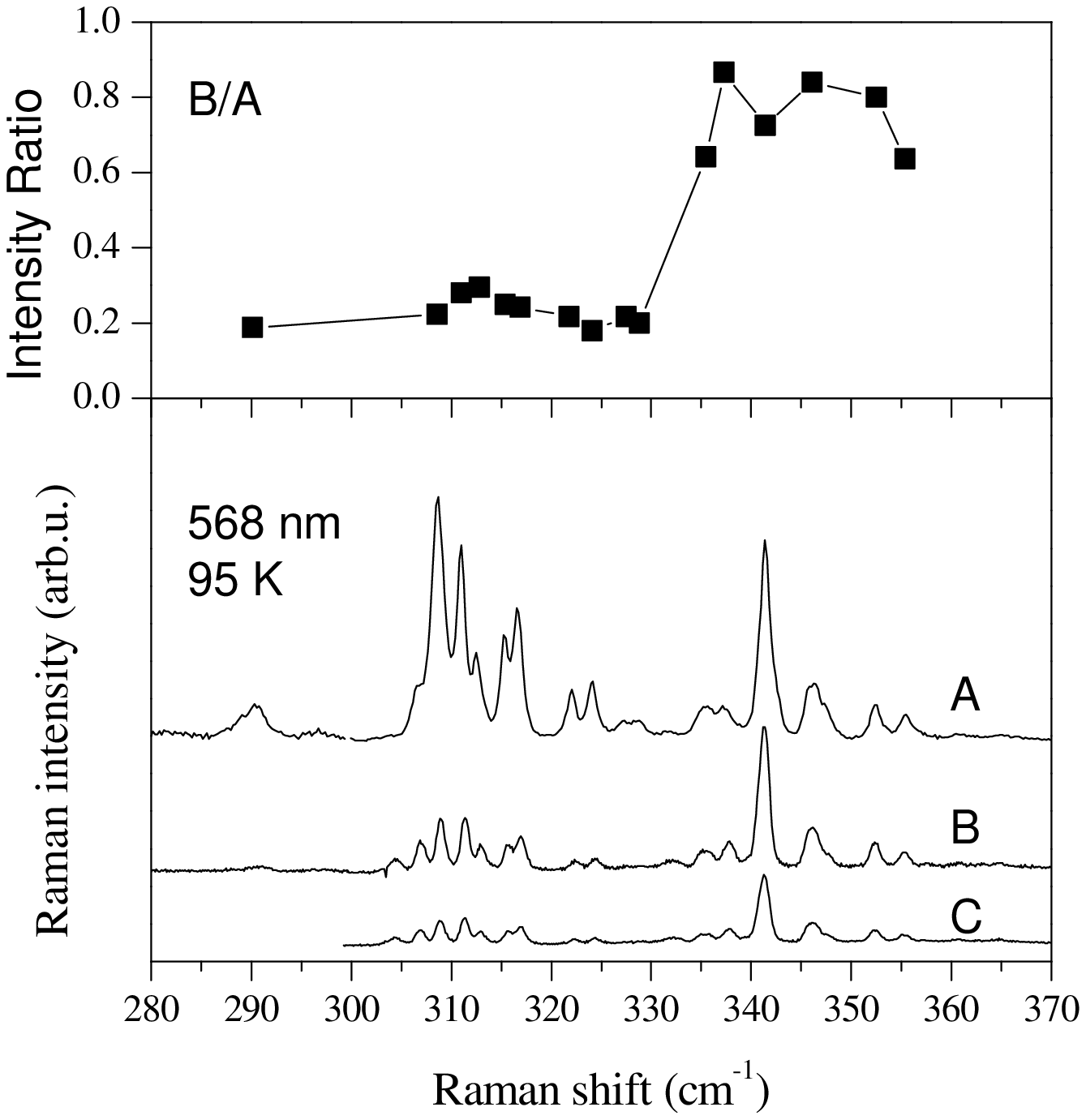}
\caption{Diameter selective growth of inner tubes in C$_{60}$ peapod based
DWCNT measured at $\protect\lambda$=568 nm laser excitation.}
\label{DWNT_growth}
\end{figure}

The inverse relationship between the RBM frequencies and tube diameters
allows a tube diameter selective study of inner tube growth. In Fig. \ref%
{DWNT_growth} we show the diameter selective growth of inner tubes. The
topmost curve, A, shows the fully developed inner tube RBM\ spectrum after a
12 h annealing at 1280$%
%TCIMACRO{\U{b0}}%
%BeginExpansion
{{}^\circ}%
%EndExpansion
$C annealing, whereas B and C correspond to 0.5 and 1 hour annealing. The
comparison reveals the more rapid development of small diameter inner tubes,
followed by the slower development of larger diameter tubes \cite%
{HolzweberKB03}. Similar diameter selective growth of inner tubes was found
by Bandow et al. \cite{BandowCPL2004}. In the latter work, it was concluded
from measurements at two laser excitations that small diameter inner tubes
form first and are subsequently transformed to larger diameter inner tubes.
This observation is somewhat different from the one reported here. The
difference might be related to the photoselective property of the Raman
experiments, as the data reported here are based on measurements on a larger
number of laser lines. The formation of smaller tubes in the beginning,
followed by the development of larger diameter tubes later on provides
useful input for the theories aimed at explaining the inner tube formation.
From computer simulation it was demonstrated that C$_{60}$ peapod based
DWCNTs are formed by Stone-Wales transformations from C$_{60}$ dimer
precursors formed at high temperature by cyclo-addition \cite%
{Tomanekprivcomm}\cite{SmalleyPRL}. The free rotation of C$_{60}$ molecules
is a prerequisite for the dimer formation as it enables the molecules to
have facing double bonds. It has been found experimentally that the
ellipsoidal shaped C$_{70}$ are present as both "standing" or "lying" peapod
configurations i.e. with the longer C$_{70}$ axis perpendicular or parallel
to the tube axis \cite{HiraharaPRB}. In small diameter tubes the lying C$%
_{70}$ configuration is preferred and the molecules have facing pentagons
and consequently no cyclo-additional double bond formation is possible in a
linear chain. For such small diameter tubes, dimers may be formed in a
canted C$_{70}$ configuration. However this structure has not been observed
experimentally. Experiments with C$_{70}$ peapod based DWCNTs have shown
that inner tubes can indeed be formed from lying C$_{70}$ peapods, which
presents a challange to the current theories \cite{Simonunpub}. Clearly, more
theoretical work is required on the formation process of inner tubes from
fullerene peapod samples.

\begin{figure}[tbp]
\centering \includegraphics[width=0.6\textwidth]{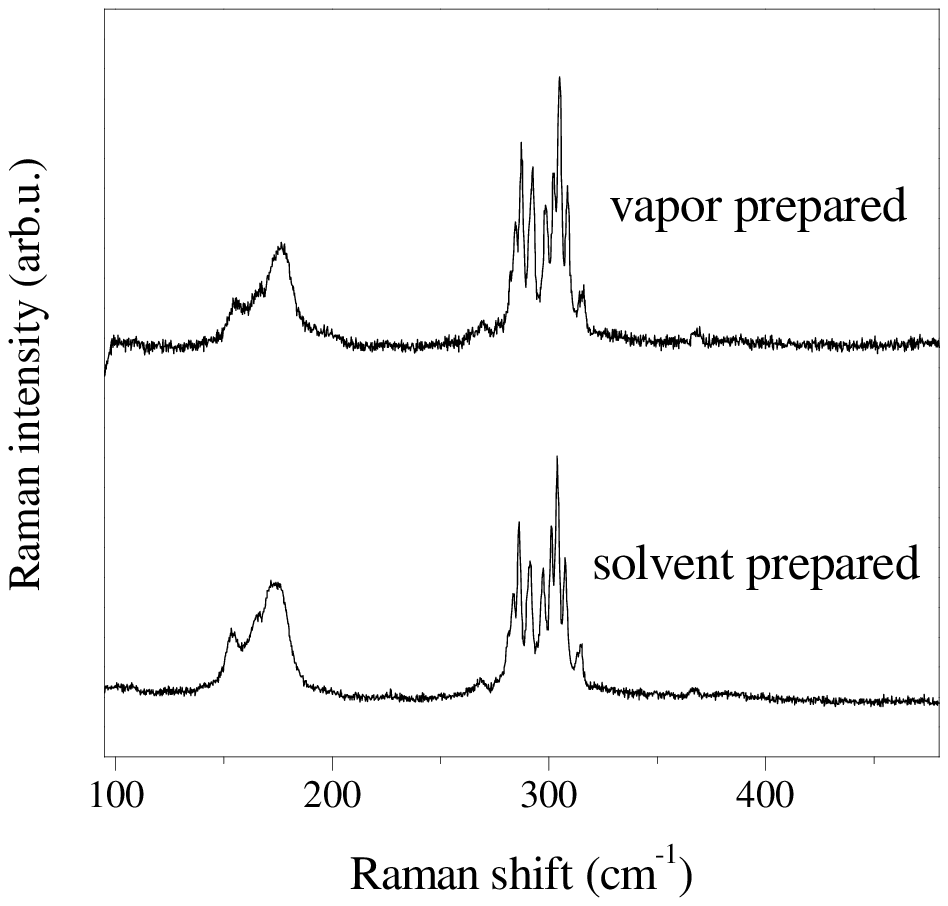}
\caption{Comparison of DWCNTs formed from vapor prepared (upper curve) and
solvent prepared (lower curve) peapod samples at $\protect\lambda$=676 nm
laser excitation and room temperature.}
\label{DWNT_solvent}
\end{figure}

Apart from the lack of full understanding of the inner tube growth, the
production process of the precursor fullerene peapods is relatively
complicated and its scaling up to larger amounts is difficult. The usual,
vapor filling, process involves the annealing of the fullerene together with
the opened SWCNT material sealed together in a quartz ampoule. Recently, a
new fullerene encapsulation technique involving the refluxing of the SWCNTs
and the fullerenes in solvents was presented \cite{SimonCPL}. In Fig. \ref%
{DWNT_solvent} we show the comparison of DWCNT samples based on vapor and
solvent filled C$_{60}$ peapods. Clearly, the yield and diameter
distribution of the inner tubes are identical in the two kinds of samples
which proves that the solvent filling method is indeed a simple alternative
to the vapor filling method. In contrast to the latter, it can be easily
scaled up to commercial quantities.

\begin{figure}[tbp]
\centering \includegraphics[width=0.6\textwidth]{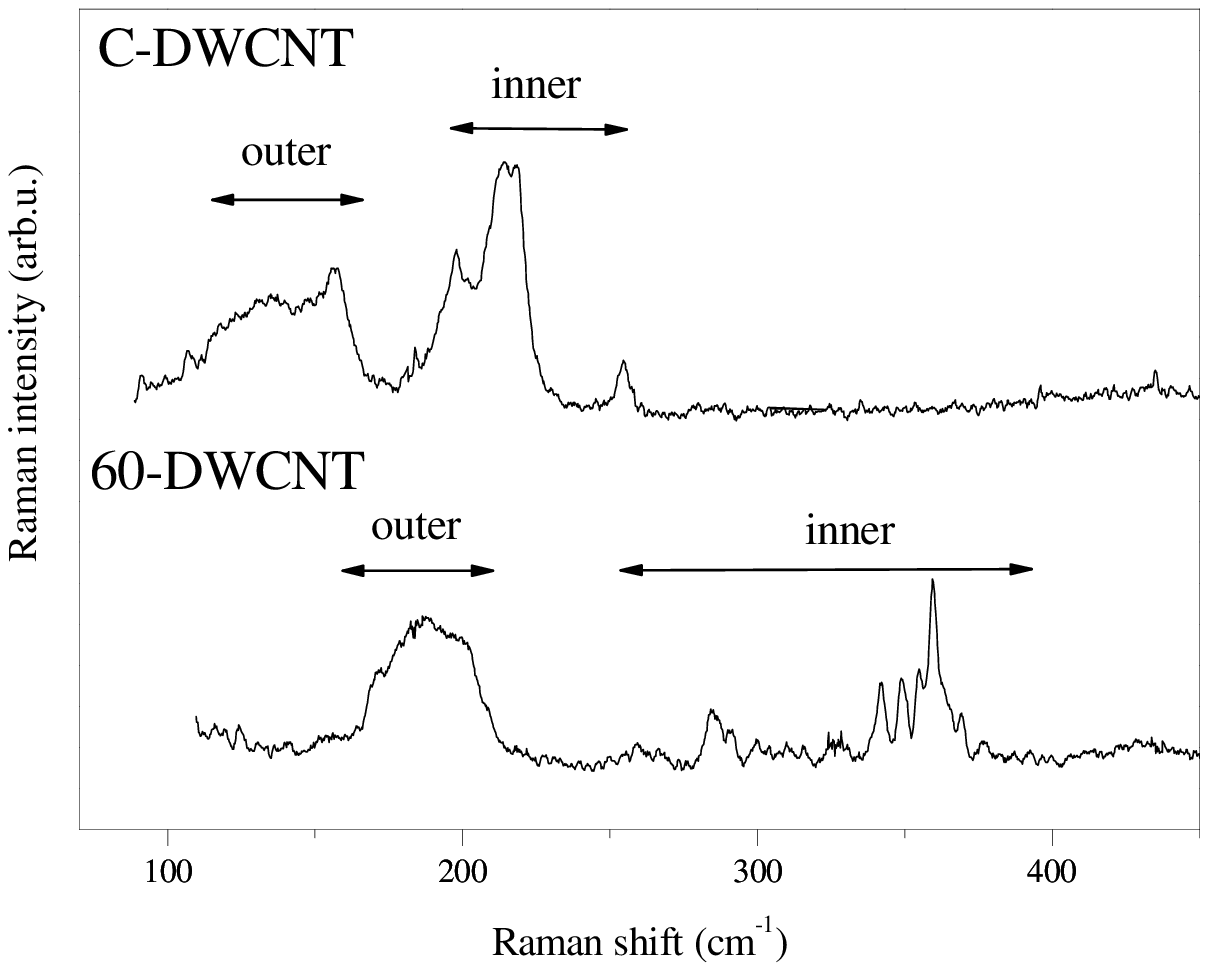}
\caption{Comparison of CVD grown (C-DWCNT) and C$_{60}$ peapod based DWCNT
(60-DWCNT) materials at $\protect\lambda$=633 nm laser excitation. Arrows
indicate the Raman shift regions where RBMs of the outer and inner tubes are
observed. The difference in inner and outer tube diameters in the two
samples is apparent.}
\label{DWNTchemC60comparison}
\end{figure}

Production of DWCNTs has been reported by direct methods \cite{HutchisonCAR}%
\cite{ChengCPL}. Such DWCNTs are characteristically different from the tubes
grown from the peapod precursors. In Fig. \ref{DWNTchemC60comparison} we
show the comparison of directly produced (C-DWCNT) and C$_{60}$ peapod based
DWCNT (60-DWCNT) materials. The C-DWCNTs were made from the catalytic
decomposition of methane and was studied in detail previously \cite%
{ChengJMatRes}. There are clear similarities and differences in the
comparision of the two kinds of DWCNT materials. Among the similarities, the
Raman spectra of both compounds show two groups of bands. The lower
frequency extending from 110-160 cm$^{-1}$ for the C-DWCNT and from 165-200
cm$^{-1}$ for the 60-DWCNT corresponds to the RBMs of the outer tubes. The
modes seen at 190-250 for the C-DWCNT and at 250-370 cm$^{-1}$ for the
60-DWCNT correspond to the RBMs of the inner tubes. The difference in the
observed RBM Raman shifts is related to the different diameters of the inner
and outer tubes in the C-DWCNT and 60-DWCNT compounds. It was found that
C-DWCNT material shown in Fig. \ref{DWNTchemC60comparison} has d$_{\mathrm{%
inner}}=$ 1.52 nm, $\sigma _{\mathrm{inner}}=$ 0.6 nm, and d$_{\mathrm{outer}}=$
2.26 nm, $\sigma _{\mathrm{outer}}=$ 0.4 nm for the mean and the variance of
the diameter distributions of the inner and outer tubes, respectively \cite%
{ChengCPL}. For the 60-DWCNT sample d$_{\mathrm{inner}}=$ 0.67 nm, $\sigma _{%
\mathrm{inner}}=$ 0.1 nm, and d$_{\mathrm{outer}}=$ 1.39 nm, $\sigma _{\mathrm{%
inner}}=$ 0.1 nm was found \cite{PfeifferPRL}. The most important difference
between the two materials is the smaller inner tube RBM linewidths for the
60-DWCNT sample. In what follows, we discuss the properties of only the
peapod based 60-DWCNT samples and will refer to these briefly as DWCNTs.

\subsection{Energy dispersive Raman studies of DWCNTs}

\subsubsection{Electronic structure of DWCNTs}

\begin{figure}[tbp]
\centering \includegraphics[width=0.6\textwidth]{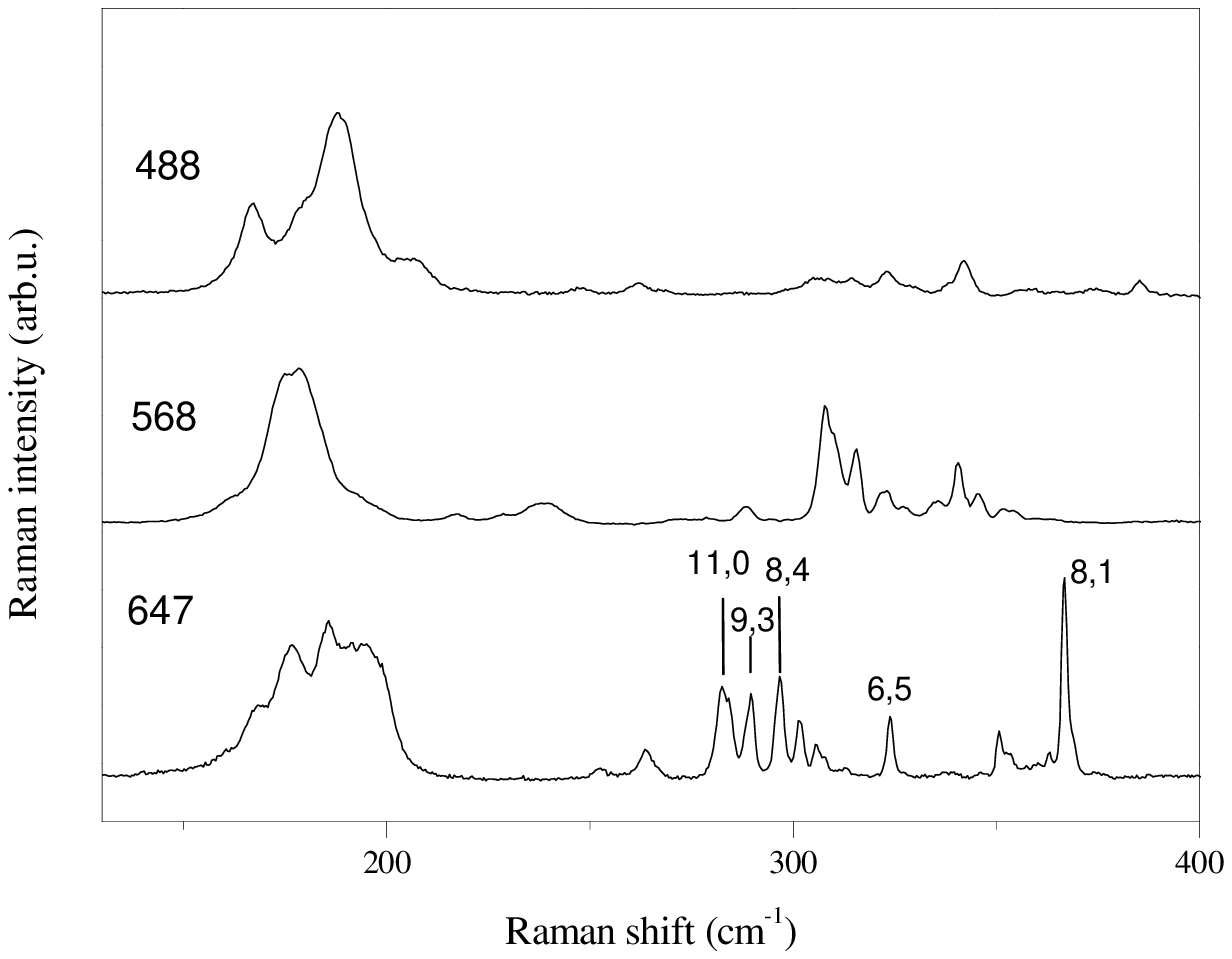}
\caption{Raman response of the RBM of a DWCNT sample when excited with
different lasers. The spectra were recorded at 90 K. The indexing for some
inner tube RBMs is shown for the $\protect\lambda$=647 nm laser energy.}
\label{DWNT_narrowing}
\end{figure}

\begin{figure}[tbp]
\centering \includegraphics[width=0.6\textwidth]{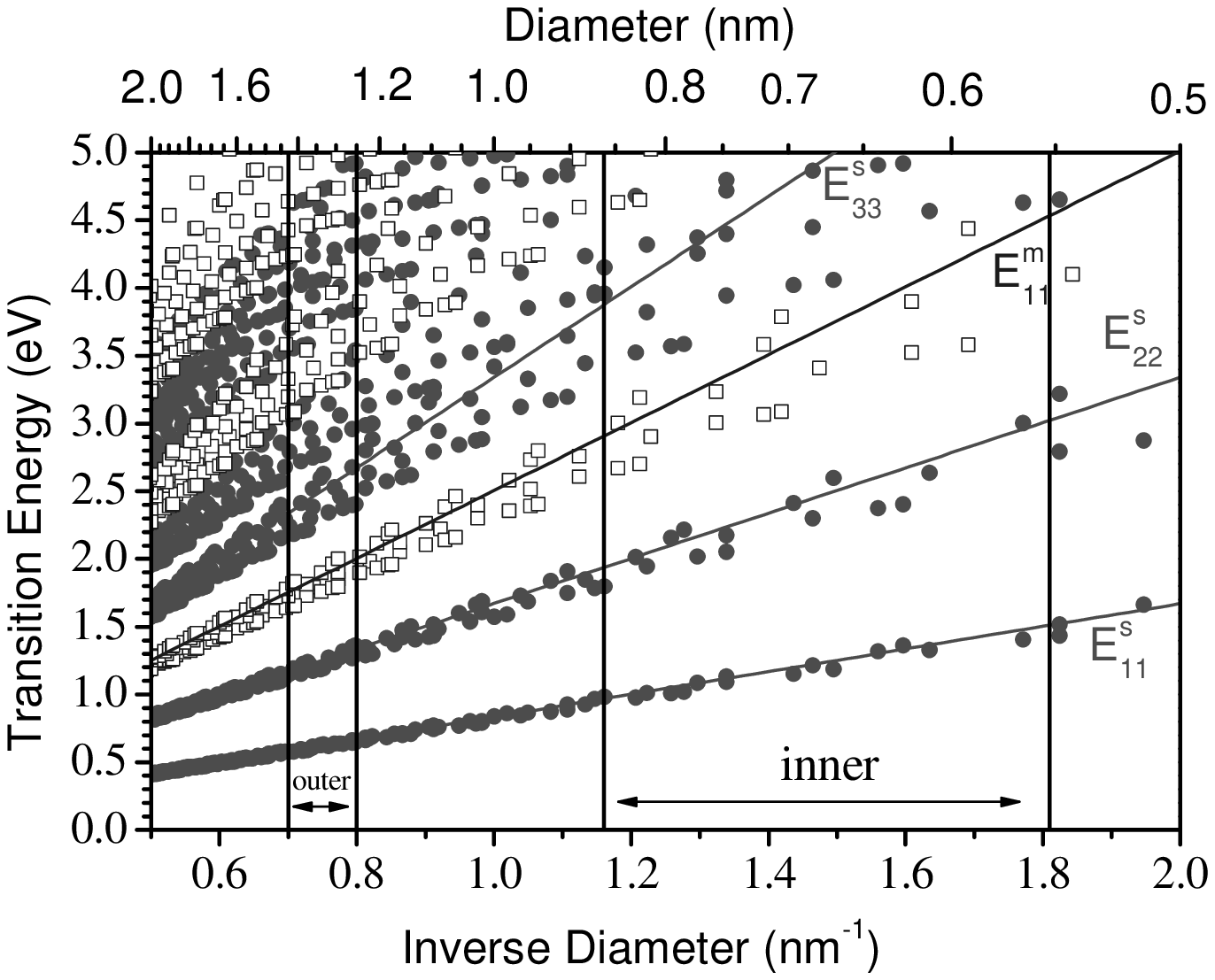}
\caption{Kataura plot with vertical lines indicating the diameter regions of
inner and outer nanotubes that are relevant for the current study.}
\label{DWNT_Kataura}
\end{figure}

The diverging behavior of the electronic density of states of SWCNTs gives
rise to a significant resonant Raman enhancement and a photoselective Raman
scattering. The photoselectivity, combined with the multi-frequency Raman
method enables a detailed study of the electronic structure of the SWCNTs.
The photoselective scattering also holds for the inner tubes of the DWCNTs.
The laser energy dependent response of the inner tube RBMs is shown in Fig. %
\ref{DWNT_narrowing}. For the red excitation RBM\ linewidths as small as 0.35
cm$^{-1}$ were observed \cite{PfeifferPRL}. This is almost an order of
magnitude smaller than reported previously on individual SWCNTs \cite%
{JorioPRL2001}. Related to the narrow linewidth of the vibrational modes,
the resonant excitation for several lines is significant, resulting in Raman
intensities almost 10 times larger than those from the outer tubes. This
implies very sharp resonances between the Van Hove singularities. This issue
is discussed in detail below. In addition to the sharp electronic
resonances, the electron-phonon coupling is enhanced for small diameter
tubes \cite{MachonKB03}, which may also contribute to the observed signal
enhancement of the inner tube vibrational modes \cite{PfeifferPRL}.

The Kataura plot that is relevant for the current study is shown in Fig. \ref%
{DWNT_Kataura}. It describes the relation between the optical transitions
and the tube diameters. It shows that when using red laser excitations (< 2 eV), 
nominally semiconducting inner tubes are expected to
be observed. The metallic inner tubes are expected to appear only for higher
energy excitations, above 2.5 eV.

\begin{figure}[tbp]
\centering \includegraphics[width=0.6\textwidth]{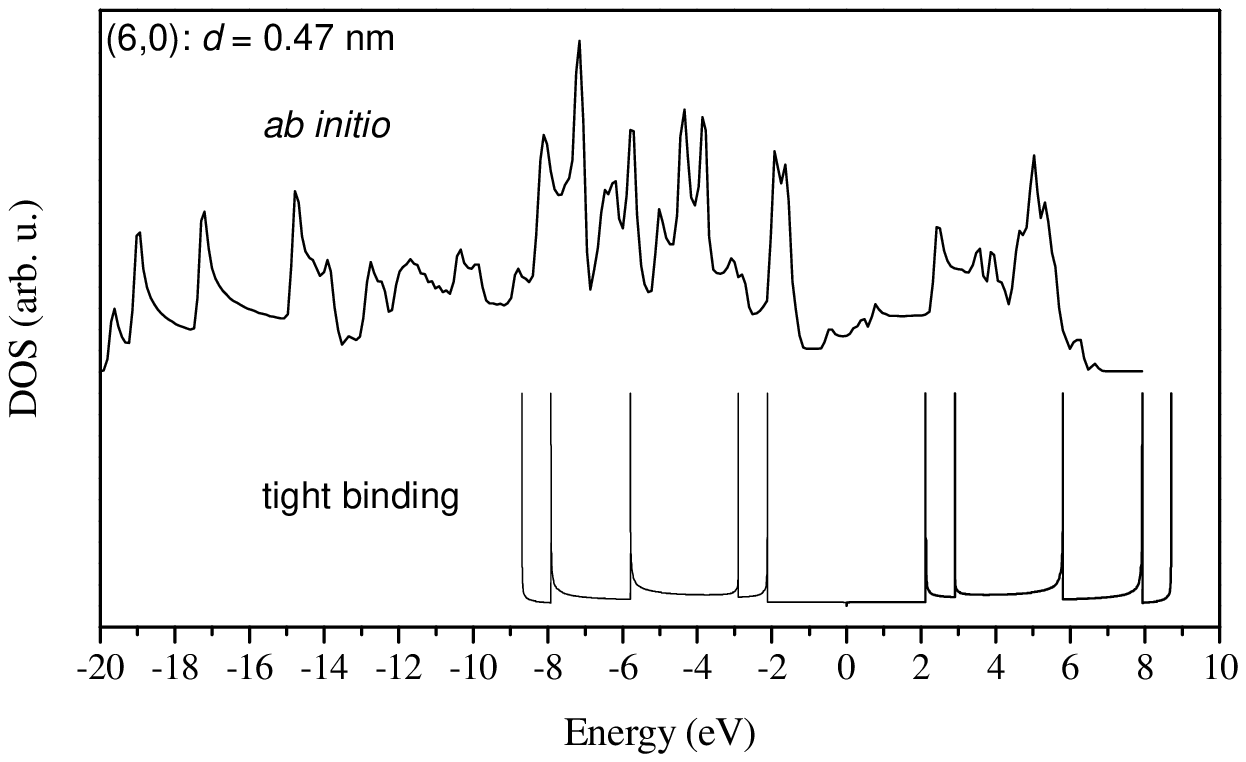}
\caption{Electronic density of states for a small diameter, metallic
nanotube using \textit{ab initio} and tight-binding techniques. The
additional structure seen in the \textit{ab initio} calculations between the
two first Van Hove singularities of the tight-binding method gives rise to
resonances at excitation energies lower than E$_{11}^{\mathrm{m}}$.}
\label{DWNT_DFT}
\end{figure}

The presence of metallic tubes, as e.g. the (9,3) inner tube at the 647 nm
laser excitation in Fig. \ref{DWNT_narrowing} is unexpected as they should
only be observable at significantly larger excitation energy when resonance
with the E$_{11}^{\mathrm{m}}$ transition occurs. This proves that smaller
energy optical transitions are present for small diameter nanotubes that are
absent in the simplest tight-binding calculations. Fig. \ref{DWNT_DFT}
compares the zone folded tight-binding and \textit{ab initio }derived
density of states (DOS) for a (6,0) metallic tube. The \textit{ab initio}
calculations show the presence of some structures between the lowest energy
optical transitions of the tight-binding model, giving rise to the
experimentally observed resonance. The observed difference between the 
\textit{ab initio} and the tight-binding theory results from the finite
curvature of the small diameter nanotubes. This curvature induces the mixing
of the $\sigma $ and the inward pointing $\pi $ orbitals.

\begin{figure}[tbp]
\centering \includegraphics[width=0.6\textwidth]{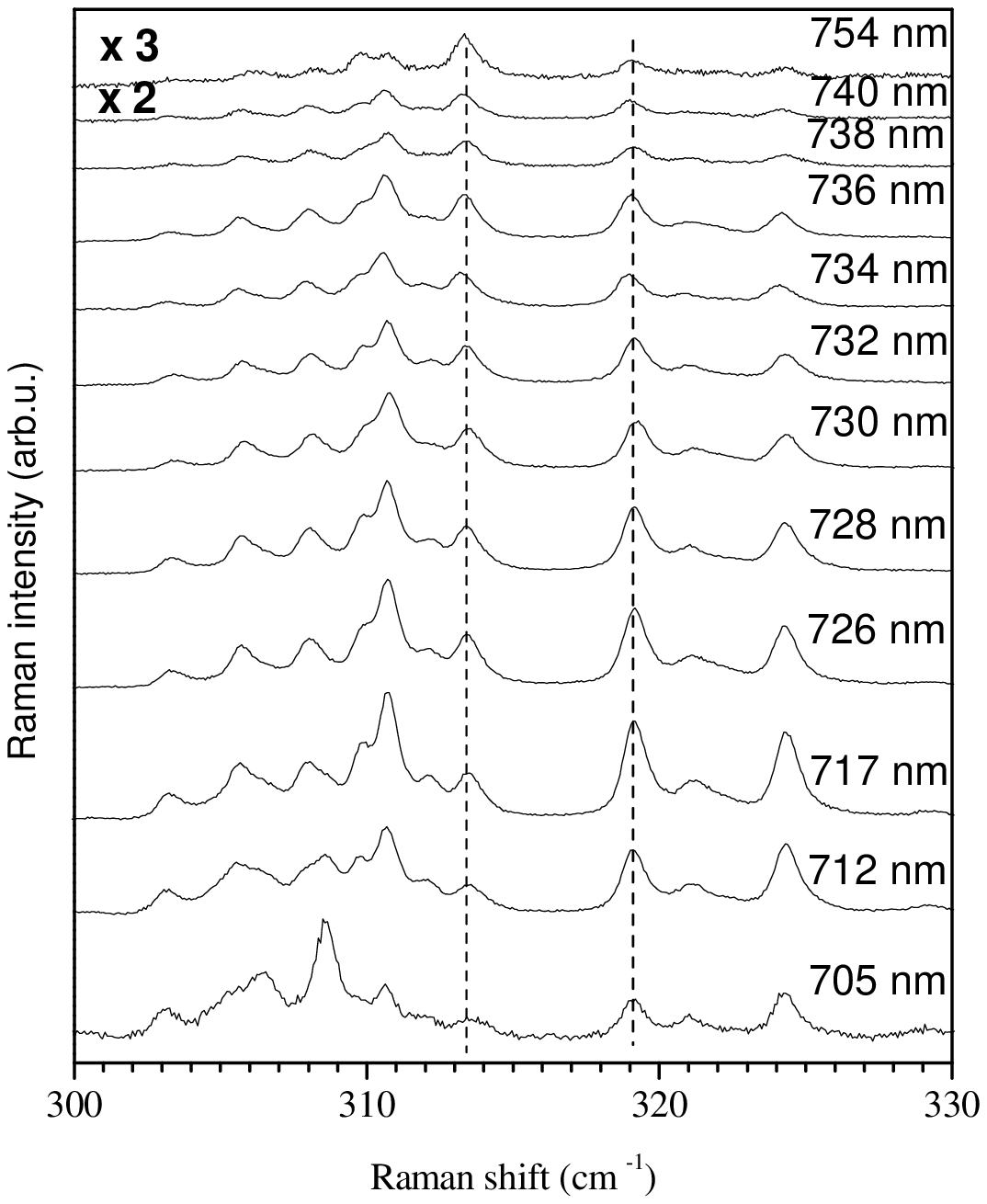}
\caption{Energy dispersive Raman spectra of the DWCNT sample in the 700-750
nm excitation energy range. Dashed lines indicate the same inner tube RBM
followed for several laser excitations. The spectra are normalized to the
incident power.}
\label{DWNT_erg_dep}
\end{figure}

The width of the Van Hove singularities of the DOS of SWCNTs is a measure of
their one-dimensional character. It can be measured from high resolution
energy dispersive Raman studies. The results in the 700-750 nm excitation
energy range are shown in Fig. \ref{DWNT_erg_dep}. Dashed lines mark the
Raman shift where the RBMs of two selected inner SWCNTs were followed for
several laser excitations. The FWHM of the energy dependent inner tube
intensities and thus the FWHM of the Van Hove singularities was found to be 60
meV \cite{SenKrichb2003}. This value is very small and reflects the one
dimensional character of the SWCNTs. However, this does not significantly
differ from values obtained from CVD grown, individual SWCNTs in a previous
study \cite{JorioERGwidth}. This shows that the high perfectness of the
inner tubes as deduced from the phonon lifetimes has no influence on the
width of the singularities in their DOS as compared to the non defect-free
outer SWCNTs. It rather supports the enhancement of electron-phonon coupling
for small diameter nanotubes as the origin for the signal enhancement. If,
however, an enhanced electron phonon coupling is present for the small
diameter tubes, one expects a range of interesting physical phenomena to
arise for the small tube such as Peierls transition \cite{Grunerbook} or
superconductivity \cite{Tinkhambook}. Indeed, superconductivity has been
observed with a critical transition temperature of T$_{c}$= 15 K in 0.4 nm
diameter SWCNTs \cite{TangSCI}. However, the relevance of such phenomena for
the inner tubes embedded in DWCNTs is not yet settled and is currently being
investigated.

\subsubsection{Splitting of the inner tube response of DWCNTs}

\begin{figure}[tbp]
\centering \includegraphics[width=0.6\textwidth]{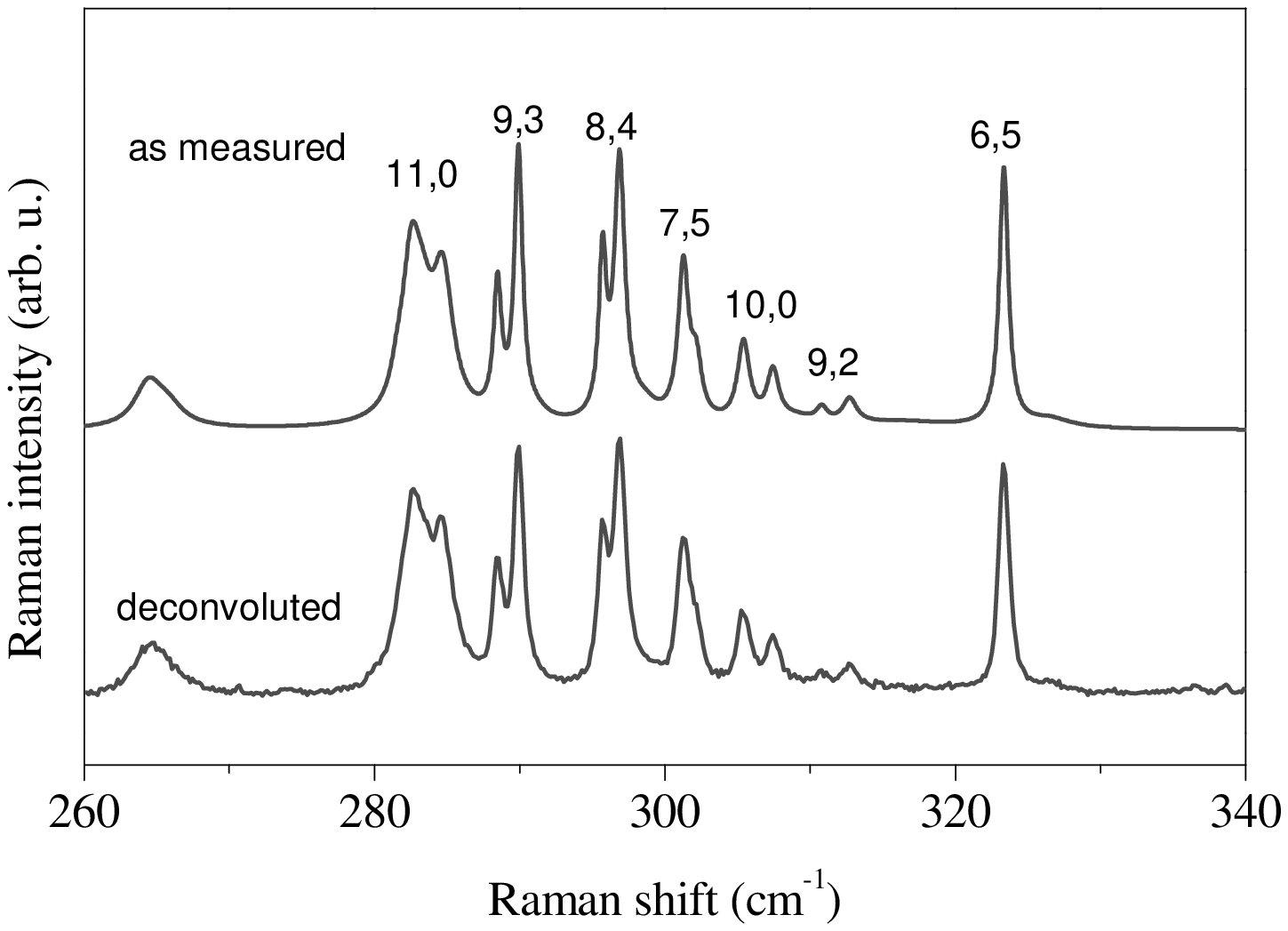}
\caption{High resolution Raman spectra of inner tube RBMs of DWCNT at $%
\protect\lambda$=647 nm laser excitation and 90 K. We also show the
corresponding spectrum after deconvolution with the resolution of the
spectrometer.}
\label{DWNT_split}
\end{figure}

In addition to the well defined number of geometrically allowed inner tubes,
a larger number of RBMs are observed. In Fig. \ref{DWNT_split} we show
spectra measured in the high resolution mode at the 647 nm laser excitation
at 90 K. The spectra after deconvolution with the resolution of the
spectrometer are also shown. The resolution of the spectrometer contributes
to an additional Gaussian broadening of the intrinsically Lorenztian RBM
lineshapes. The width of the Gaussian was measured from the response of our
apparatus to the exciting laser and was found to be 0.4-0.7 cm$^{-1}$
depending on the laser energy. The presence of additional, split
componenents is apparent in Fig. \ref{DWNT_split} for some tubes. Some RBMs
split into even 3 and more componenents. This splitting is a natural
consequence of the different number of geometrically allowed inner and outer
tubes and is related to the interaction between the two shells of the
DWCNTs. As the diameters of both the inner and outer tubes are discrete
sequences, some inner tubes can be grown in outer tubes with different
diameters. Then the difference in inner-outer tube wall disctance gives rise
to a different interaction that causes the observed splitting of the lines.
A rough estimate yields that for the DWCNTs studied here, 40 geometrically
allowed outer tubes accommodate 20 geometrically allowed inner tubes. As a
consequence, on average splitting into two components is expected. This
estimate, however, does not account for the magnitude of the splitting that
is currently being further investigated both experimentally and
theoretically \cite{splittingpaper}.

\begin{figure}[tbp]
\centering \includegraphics[width=0.6\textwidth]{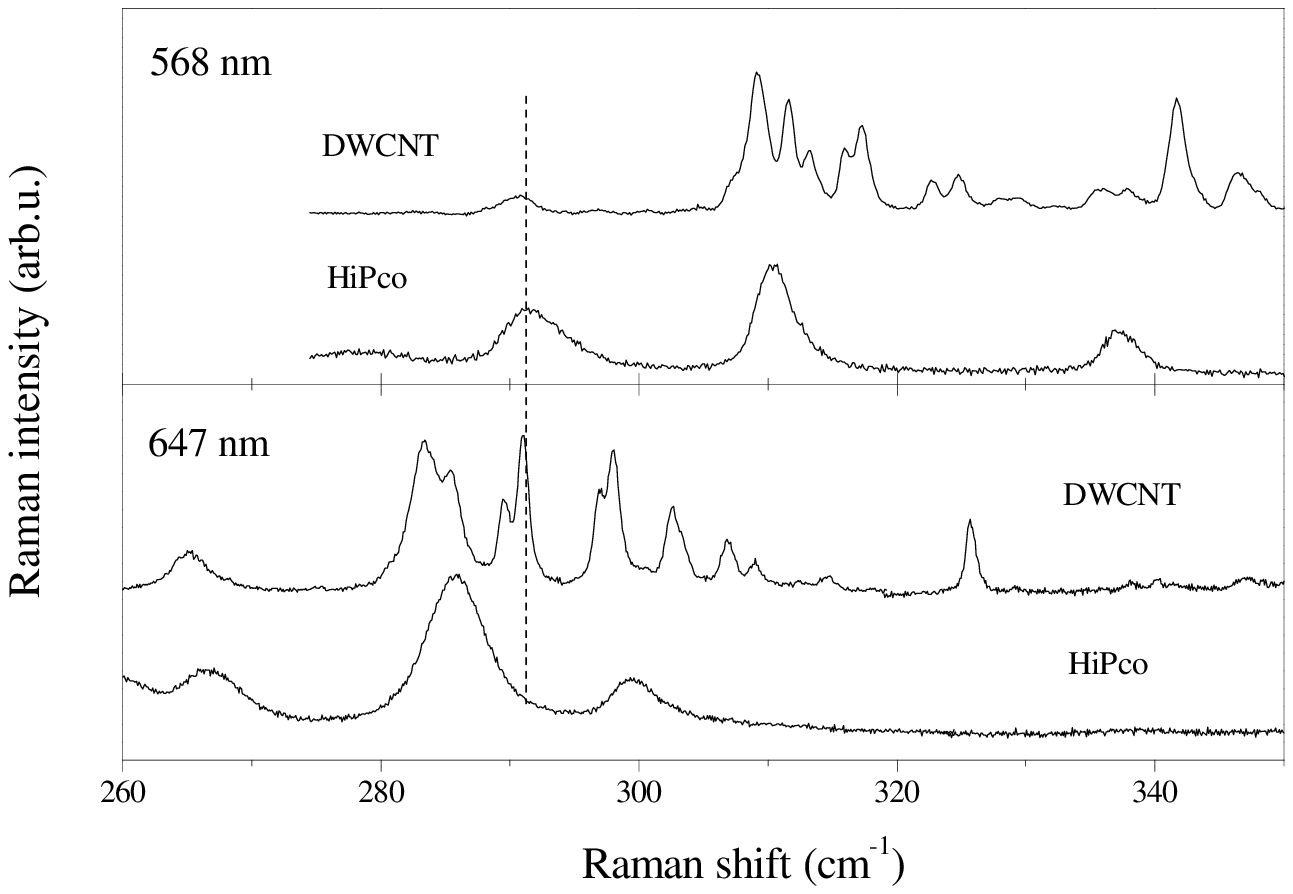}
\caption{Comparison of the inner tube RBMs of a DWCNT sample and the RBMs of
a sample prepared by the HiPco process at $\protect\lambda$=568 nm and 647
nm laser energy. Both samples were measured at 90 K in the high resolution
mode.}
\label{DWNT_HIPCO}
\end{figure}

In addition to the two-wall related splitting of the RBMs of inner tubes,
further peculiarities can be observed. In Fig. \ref{DWNT_HIPCO} we compare
the inner tube RBMs of a DWCNT\ sample with the RBMs of a small diameter
SWCNT sample. The latter was a HiPco sample with a mean diameter and a
variance of $d$ =1.05 nm and $\sigma $=0.15 nm, respectively \cite%
{KukoveczEPJB}. The figure shows this comparison for 568 nm and 647 nm laser
excitation. The larger number of RBMs in the DWCNT sample as compared to the
HiPco sample and the absence of splitting for the RBMs in the HiPco sample
is observed. Broader RBM lines are observed in the HiPco sample, however
this would not limit the observation of the splitting. As discussed above,
the splitting is related to the two-shell nature of the DWCNT samples and
thus its absence is natural in the HiPco sample. However, the absence of
some RBMs corresponding to geometrically allowed tubes in the HiPco sample
that are osberved in the inner tube DWCNT spectrum is intriguing. The
absence of geometrically allowed SWCNTs or the smaller number of optical
transitions in the HiPco samples may explain for our observation. The dashed
line in Fig. \ref{DWNT_HIPCO} shows an example for a tube RBM that is absent
at 647 nm excitation from the HiPco spectrum, however appears at 568 nm
excitation. Similar behavior was observed for other, missing HiPco RBM
modes, i.e. the RBM modes of the SWCNTs of the HiPco samples are also
present, altough at much less number of laser lines. This clearly shows,
that all the geometrically allowed SWCNTs are present both among the inner
tubes and also in the HiPco sample, however, there is a significantly larger
number of optical transitions for the inner tubes of the DWCNT sample. This
is most probably related to a yet unexplained intricate interplay between
the two shells of the DWCNT samples and calls for theoretical work on this
issue. In addition to the larger number of lines observed in the inner tube
RBM spectrum of DWCNTs, a slight downshift ranging from 2-3 cm$^{-1}$ is
also observed. This downshift is related to the different environment for an
inner tube of the DWCNT and for an SWCNT in a HiPco sample. The earlier is
surrounded by an outer tube, whereas the latter is embedded in bundles.
These interactions give rise to a different value for the C$_{2}$ constant
of the RBM mode frequencies. The broader linewidths observed in the HiPco
sample underlines the highly perfect nature of the inner nanotubes.

\begin{figure}[tbp]
\centering \includegraphics[width=0.6\textwidth]{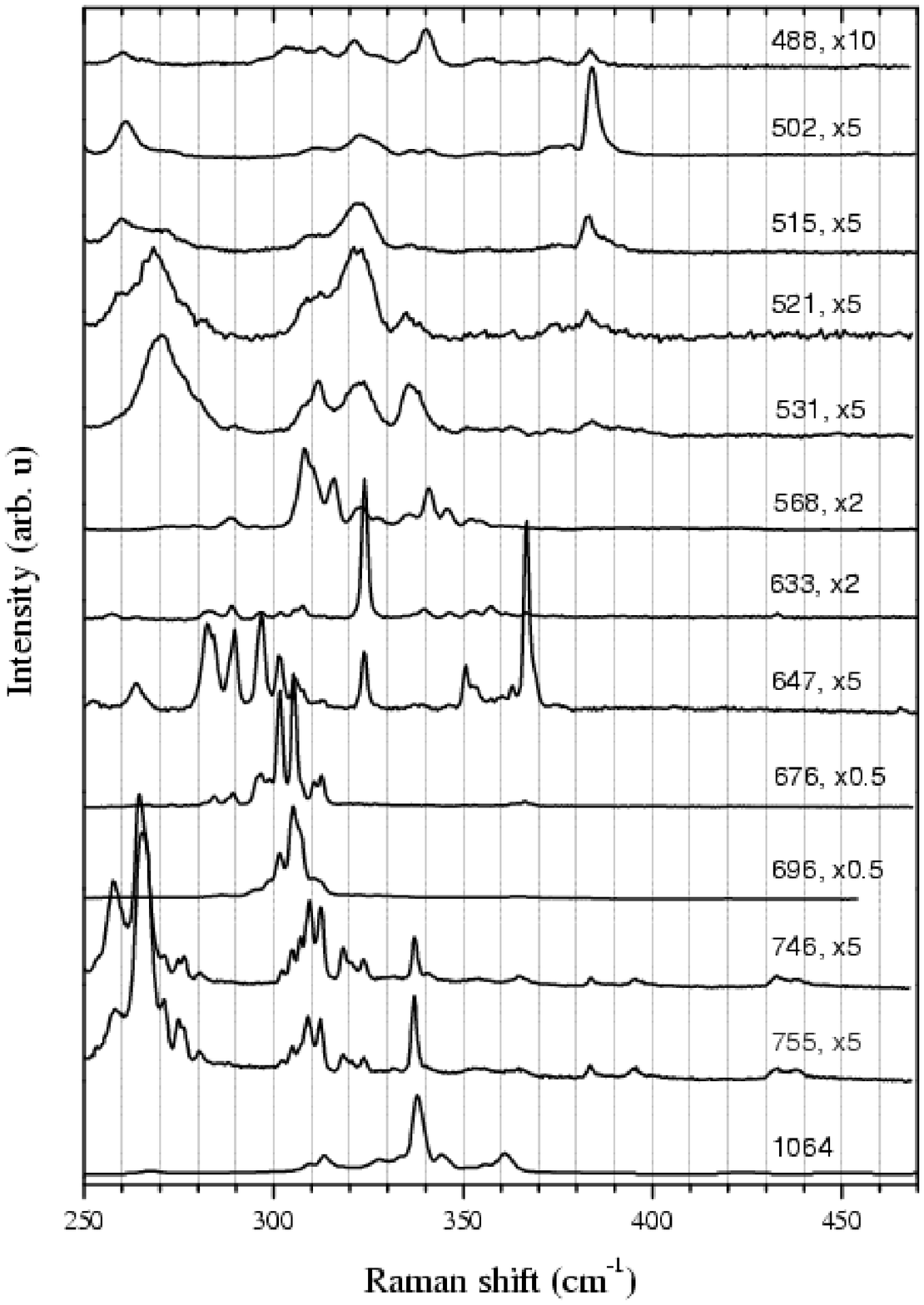}
\caption{DWCNT RBM spectra measured at 90 K for several laser excitations. }
\label{DWNT_summary}
\end{figure}

\subsubsection{Chiral index assigment for inner tubes}

The reciprocal relation between the RBM frequencies and the tube diameters
contributes to a significant spectral spread for the observed inner tube
RBMs. This, together with the narrow linewidths observed for the inner tubes
allows the accurate ($n,m$) indexing of the tubes. In Fig. \ref{DWNT_summary},
the inner tube RBMs measured at several laser lines are shown. The
observation of a large number of well defined RBMs allows the determination
of RBM shifts corresponding to distinct inner tubes. The Raman shifts of
observable inner tubes are summarized in Table. \ref{DWNT_shift_table},
together with the assigned chiral vectors, and tube diameters calculated
with DFT methods. It was found that the diameters of small tubes, based on
the lattice constant of graphene gives significant deviations as compared to
the DFT calculated tube diameters \cite{KrambergerPRB}.

\begin{figure}[tbp]
\centering \includegraphics[width=0.6\textwidth]{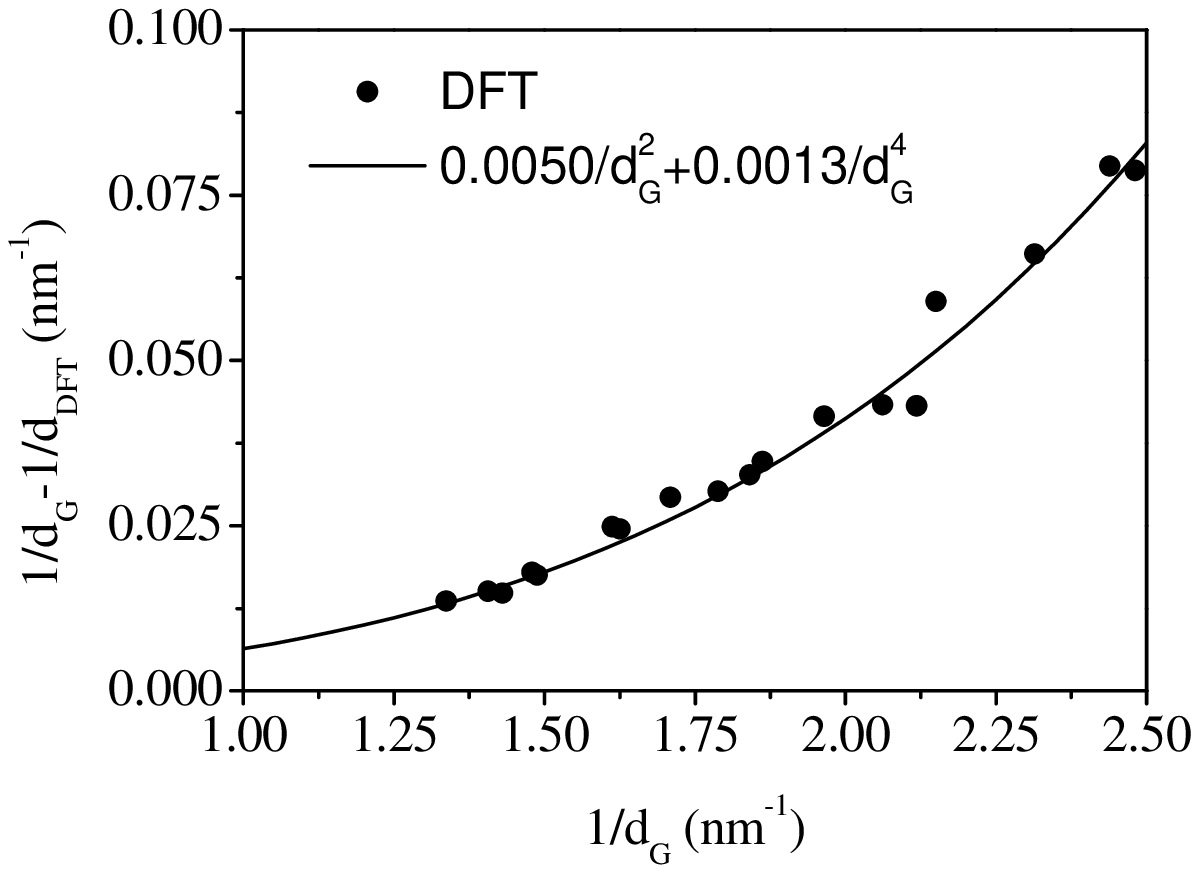}
\caption{Difference between graphene and DFT derived inverse tube diameters.
The solid line is a polynomial interpolation.}
\label{DWNT_DFT_diams}
\end{figure}

\begin{figure}[tbp]
\centering \includegraphics[width=0.6\textwidth]{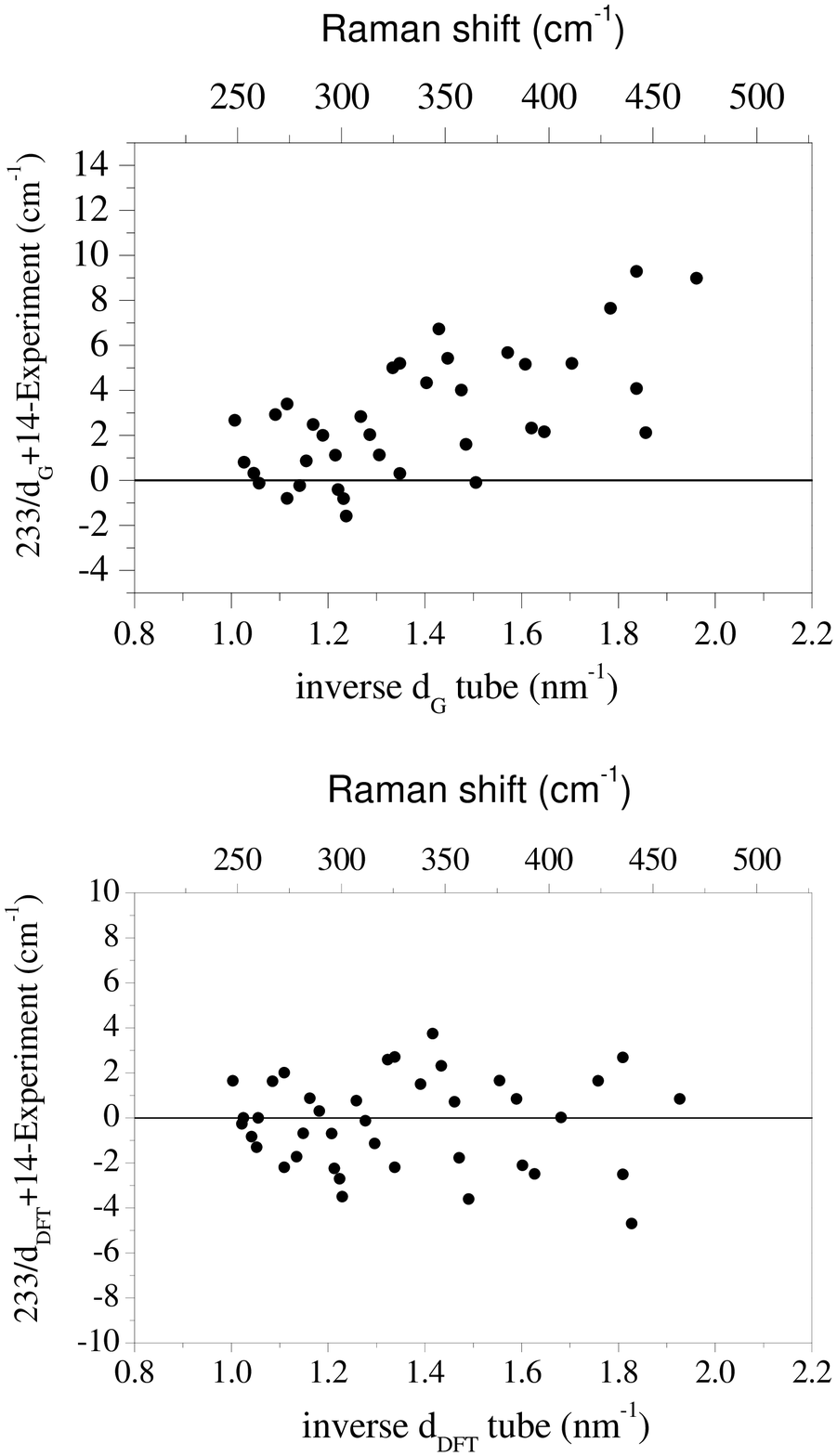}
\caption{Difference between the calculated and measured RBM Raman shifts
when the graphene derived (upper panel) and DFT optimized (lower panel) tube
diameters were used.}
\label{DWNT_shift_diff}
\end{figure}

\begin{table}[tbp]
\caption{Inner tube RBM frequencies and tube diameters: (1) center of
gravity line position of the RBM averaged from different laser excitations,
(2) CNT chiral indices, (3) interpolated DFT determined tube diameters, (4)
first and second tight-binding optical transition energies with $\protect%
\gamma_{0}$= 2.9 eV, (5) theoretical RBM frequencies at the best fit with $%
C_{1}$=233 cm$^{-1}$ nm and $C_{2}$= 14 cm$^{-1}$. ("n.i.": not identified),
from \protect\cite{KrambergerPRB}.}
\label{DWNT_shift_table}\centering \renewcommand{\arraystretch}{1.2} %
\setlength\tabcolsep{5pt} 
\begin{tabular}{@{}llllll@{}lll@{}lll}
\hline
\noalign{\smallskip} $\nu_{\mathrm{RBM}}$ expt. (cm$^{-1}$) & Chirality (n,m)
& d$_{\mathrm{RBM}}$ (nm) & E$_{11}$/E$_{22}$ (eV) & $\nu _{\mathrm{RBM}}$theor.
(cm$^{-1}$) &  &  &  &  &  &  &  \\ \hline
\noalign{\smallskip} 246.1 & (11,3) & 0.997 & 0.84/1.59 & 247.8 &  &  &  & 
&  &  &  \\ 
252.4 & (12,1) & 0.978 & 0.86/1.60 & 252.1 &  &  &  &  &  &  &  \\ 
n.i. & (10,4) & 0.975 & 2.36/4.21 & 252.9 &  &  &  &  &  &  &  \\ 
257.6 & (9,5) & 0.960 & 0.84/1.73 & 256.8 &  &  &  &  &  &  &  \\ 
260.5 & (8,6) & 0.950 & 0.87/1.69 & 259.2 &  &  &  &  &  &  &  \\ 
n.i. & (11,2) & 0.947 & 2.40/4.24 & 260 &  &  &  &  &  &  &  \\ 
n.i. & (7,7) & 0.947 & 2.52/4.54 & 260 &  &  &  &  &  &  &  \\ 
n.i. & (12,0) & 0.937 & 2.40/4.25 & 262.6 &  &  &  &  &  &  &  \\ 
265.3 & (10,3) & 0.921 & 0.87/1.84 & 267 &  &  &  &  &  &  &  \\ 
270.5 & (11,1) & 0.901 & 0.85/1.91 & 272.5 &  &  &  &  &  &  &  \\ 
274.7 & (9,4) & 0.901 & 0.82/1.75 & 272.5 &  &  &  &  &  &  &  \\ 
n.i. & (8,5) & 0.888 & 2.61/4.60 & 276.5 &  &  &  &  &  &  &  \\ 
280.2 & (7,6) & 0.881 & 0.93/1.85 & 278.5 &  &  &  &  &  &  &  \\ 
282.3 & (10,2) & 0.871 & 0.97/1.78 & 281.6 &  &  &  &  &  &  &  \\ 
284.0 & (11,0) & 0.860 & 0.98/1.80 & 284.9 &  &  &  &  &  &  &  \\ 
289.1 & (9,3) & 0.846 & 2.67/4.63 & 289.4 &  &  &  &  &  &  &  \\ 
296.1 & (8,4) & 0.828 & 0.98/2.00 & 295.4 &  &  &  &  &  &  &  \\ 
298.9 & (10,1) & 0.824 & 2.70/4.65 & 296.7 &  &  &  &  &  &  &  \\ 
301.9 & (7,5) & 0.817 & 1.01/1.95 & 299.2 &  &  &  &  &  &  &  \\ 
304.0 & (6,6) & 0.813 & 2.90/5.02 & 300.5 &  &  &  &  &  &  &  \\ 
306.5 & (9,2) & 0.794 & 1.01/2.15 & 307.3 &  &  &  &  &  &  &  \\ 
311.7 & (10,0) & 0.783 & 1.02/2.22 & 311.6 &  &  &  &  &  &  &  \\ 
317.2 & (8,3) & 0.771 & 1.09/2.02 & 316.1 &  &  &  &  &  &  &  \\ 
319.8 & (7,4) & 0.756 & 3.00/5.07 & 322.5 &  &  &  &  &  &  &  \\ 
323.0 & (9,1) & 0.748 & 1.13/2.05 & 325.8 &  &  &  &  &  &  &  \\ 
327.9 & (6,5) & 0.748 & 1.09/2.17 & 325.8 &  &  &  &  &  &  &  \\ 
336.7 & (8,2) & 0.719 & 3.06/5.09 & 338.3 &  &  &  &  &  &  &  \\ 
340.3 & (9,0) & 0.706 & 3.08/5.10 & 344.1 &  &  &  &  &  &  &  \\ 
345.8 & (7,3) & 0.697 & 1.15/2.41 & 348.2 &  &  &  &  &  &  &  \\ 
353.8 & (6,4) & 0.684 & 1.21/2.30 & 354.6 &  &  &  &  &  &  &  \\ 
358.5 & (5,5) & 0.680 & 3.37/5.52 & 356.8 &  &  &  &  &  &  &  \\ 
364.9 & (8,1) & 0.671 & 1.19/2.60 & 361.4 &  &  &  &  &  &  &  \\ 
374.5 & (7,2) & 0.643 & 1.32/2.38 & 376.3 &  &  &  &  &  &  &  \\ 
383.5 & (8,0) & 0.629 & 1.36/2.40 & 384.4 &  &  &  &  &  &  &  \\ 
389.3 & (6,3) & 0.624 & 3.52/5.54 & 387.3 &  &  &  &  &  &  &  \\ 
395.6 & (5,4) & 0.615 & 1.33/2.63 & 393.2 &  &  &  &  &  &  &  \\ 
405.8 & (7,1) & 0.595 & 3.58/5.55 & 405.9 &  &  &  &  &  &  &  \\ 
422.0 & (6,2) & 0.569 & 1.40/3.00 & 423.8 &  &  &  &  &  &  &  \\ 
432.9 & (7,0) & 0.553 & 1.43/3.22 & 435.7 &  &  &  &  &  &  &  \\ 
438.1 & (5,3) & 0.553 & 1.52/2.79 & 435.7 &  &  &  &  &  &  &  \\ 
444.5 & (4,4) & 0.547 & 4.10/5.80 & 439.9 &  &  &  &  &  &  &  \\ 
462.1 & (6,1) & 0.519 & 1.66/2.87 & 463.1 &  &  &  &  &  &  &  \\ \hline
&  &  &  &  &  &  &  &  &  &  & 
\end{tabular}%
\end{table}

In Fig. \ref{DWNT_DFT_diams} we compare the inverse tube diameters
calculated from the lattice constant of graphene and from a DFT calculation.
A simple interpolation could be established: 1/$d_{\mathrm{DFT}}=$ 1/$d_{\mathrm{%
G}}$ - $\left( 0.0050/d_{\mathrm{G}}^{2}+0.0013/d_{\mathrm{G}}^{4}\right) $,
where $d_{\mathrm{DFT}}$ and $d_{\mathrm{G}}$ are the DFT and the graphene
derived tube diameters, respectively, and $d_{\mathrm{G}}^{{}}$ is expressed
from the chiral indices as: $d_{\mathrm{G}}^{{}}=0.141\sqrt{3\left(
m^{2}+n^{2}+mn\right) }/\pi $. The $C_{1}$ and $C_{2}$ constants of the RBM
frequencies $\nu _{\mathrm{RBM}}$=$C_{\mathrm{1}}/d_{\mathrm{DFT}}+C_{2}$ were
determined from a linear regression and $C_{1}$=233 cm$^{-1}$ nm and $C_{2}$%
=14 cm$^{-1}$ were found. Fig. \ref{DWNT_shift_diff} shows that no linear
relationship could be established between the RBM frequencies and the
graphene derived inverse tube diameters. However, a very reliable fit is
obtained with small discrepancies between calculated and measured RBM
frequency values when the DFT optimized tube diameters were used. It
establishes that the reciprocal relationship between the RBM frequencies and
the tube diameters, when the latter is properly calculated, is valid down to
the smallest observable inner tubes. The value determined for the $C_{1}$
constant is not restricted to small diameter tubes only, however studies on
larger diameter tubes have been lacking the precision that could be obtained
from the inner tube analysis of DWCNTs. Therefore a direct comparison with
previously determined values can not be performed. It has been pointed out
above that the different environment for an inner tube and for an SWCNT in a
bundle prevents a comparison of the respective C$_{2}$ parameters.

In conclusion, the Raman studies of peapod based DWCNT materials have been
reviewed. It was shown that this material has an unprecedently high
perfectness related to the growth in the catalyst free environment. The
growth mechanism is not fully understood and some alternative mechanisms
have been discussed. A method is presented that enables the industrial
scaling up of the DWCNT production. The properties of the DWCNT have been
compared with a similar diameter SWCNT material, a HiPco sample. The
geometrically allowed small diameter tubes are present in both materials.
However, due to the smaller number of optical transitions in the HiPco
sample, a smaller number of RBM lines are observed. A splitting of the
geometrically allowed inner tube RBMs in the DWCNT sample was observed. It
was explained by the different shell-shell distance of inner-outer tube
pairs with varying diameters. The sharp appearance of the inner tube RBMs
and their relatively larger spectral spread allowed a chiral index
assignment for a broad spectral range and Raman shifts. It was found that a
linear relation could be established between the RBM frequencies and DFT
determined tube diameters. The empirical constants relating the RBM
frequencies and the tube diameters have been refined. Nevertheless, a direct
comparison with values determined on free-standing or bundled SWCNTs is not
straigthforward.

\section{Acknowledgements}

The authors gratefully acknowledge J. K\"{u}rti, and V. Z\'{o}lyomi for many
fruitful discussions and for the DFT calculations. H. Kataura is
acknowledged for providing some of the C$_{60}$ peapod materials. We thank
Hui-Ming Cheng for providing the CVD grown DWCNT sample. The authors
gratefully acknowledge J. Bernardi for the TEM micrographs, Th. Pichler and 
\'{A}. Kukovecz for their contributions to the Raman experiments. This work
was supported by the Austrian Science Funds (FWF) project Nr. 14893 and by
the EU projects NANOTEMP BIN2-2001-00580 and PATONN Marie-Curie
MEIF-CT-2003-501099.

%\printindex

\end{document}